\documentclass[preprintnumbers,nofootinbib,prc,superscriptaddress]{revtex4-1}

\usepackage{graphicx}
\usepackage{amssymb}
\usepackage{amsmath}
\usepackage{epstopdf}
\usepackage{placeins}
\usepackage{bm}

\begin{document}
    
\title{Recoil-Order and Radiative Corrections to the aCORN Experiment}
    
\author{F.~E.~Wietfeldt}
\affiliation{Department of Physics and Engineering Physics, Tulane University, New Orleans, LA 70118}
\author{W.~A.~Byron}\thanks{current address: Dept. of Physics, University of Washington, Seattle, WA 98195, USA}
\affiliation{Department of Physics and Engineering Physics, Tulane University, New Orleans, LA 70118}
\author{B.~Collett}
\affiliation{Physics Department, Hamilton College, Clinton, NY 13323}
\author{M.~S.~Dewey}
\affiliation{National Institute of Standards and Technology, Gaithersburg, MD 20899, USA}
\author{T.~R.~Gentile}
\affiliation{National Institute of Standards and Technology, Gaithersburg, MD 20899, USA}
\author{F.~Gl\"{u}ck}
\affiliation{Karlsruhe Institute of Technology, IAP, 76021 Karlsruhe, POB 3640, Germany}
\author{M.~T.~Hassan}
\affiliation{Department of Physics and Engineering Physics, Tulane University, New Orleans, LA 70118}
\author{G.~L.~Jones}
\affiliation{Physics Department, Hamilton College, Clinton, NY 13323}
\author{A.~Komives}
\affiliation{Department of Physics and Astronomy, DePauw University, Greencastle, IN 46135}
\author{J.~S.~Nico}
\affiliation{National Institute of Standards and Technology, Gaithersburg, MD 20899, USA}
\author{E.~J.~Stephenson}
\affiliation{CEEM, Indiana University, Bloomington, IN 47408}

\date{\today}

\begin{abstract}
The aCORN experiment measures the electron-antineutrino $a$-coefficient in free neutron decay. We update the previous aCORN results \cite{Dar17,Has21} to include radiative and recoil corrections to first order, and discuss a key issue in the comparison of results from different $a$-coefficient experimental methods when these effects are considered.  The corrected combined result is $\overline{a} = -0.10779 \pm 0.00125\, ({\rm stat}) \pm 0.00133\, ({\rm sys})$, averaged over the full Fermi neutron beta spectrum. The corresponding corrected result for the ratio of weak coupling constants $\lambda = G_A/G_V$ is $\lambda = -1.2712 \pm 0.0061$. This improves agreement with previous $a$-coefficient experiments, in particular the 2020 aSPECT result \cite{Bec20}.
\end{abstract}

\maketitle

\section{Introduction}
\label{S:Intro}
Beta decay of the free neutron $n \rightarrow p + e + \overline{\nu_e}$  is the simplest nuclear beta decay system and as such it is a useful laboratory for precise studies of the weak nuclear force and low energy tests of the Standard Model of particle physics (SM). The key experimental features of neutron decay are given by the formula of Jackson, Treiman, and
Wyld \cite{JTW}, the neutron decay probability as a function of the emitted electron ($e$) and antineutrino ($\nu$) momentum vectors ${\bf p_e}$, ${\bf p_{\nu}}$ 
and total energies $E_e$, $E_\nu$, and the neutron spin polarization vector ${\bf P}$
\begin{equation}
\label{E:JTWeqn}
d^3\Gamma \propto \frac{1}{\tau_n}E_e |{\bf p_e}| (E_e^{\rm max}-E_e)^2 \Bigg[ 1 + a\frac{{\bf p_e}\cdot {\bf p_{\nu}}}{E_e E_\nu} + b\frac{m_e}{E_e} 
+ {\bf P}\cdot \left( A\frac{\bf p_e}{E_e} + B\frac{{\bf p_{\nu}}}{E_\nu} + D\frac{({\bf p_e}\times {\bf p_{\nu}})}{E_e E_\nu} \right) \Bigg] dE_e d\Omega_e d\Omega_{\nu}
\end{equation}
where $\tau_n$ is the neutron lifetime and $E_e^{\rm max}$ = 1292 keV is the endpoint energy. The parameters $a$, $A$, $B$, and $D$ are 
experimentally measured correlation coefficients. In the SM these depend on $\lambda = G_A / G_V$, the ratio of the weak axial vector ($G_A$) and vector ($G_V$) couplings between free protons and neutrons. The Fierz interference parameter $b$ is zero in the SM; it would be generated by the presence of scalar or tensor weak currents. The aCORN experiment measures the electron-antineutrino correlation $a$-coefficient.  Precise experimental values of neutron decay parameters are important in cosmology, astrophysics, and neutrino detection \cite{Dub91,Bar05,Cyb16,Dub21}. $G_V$ is used to determine the element $V_{ud}$ of the Cabibbo–Kobayashi–Maskawa (CKM) quark mixing matrix and test its unitarity \cite{Har15}. The currently most precise determinations of $\lambda$ are from the UCNA \cite{Bro18} and PERKEO III \cite{Mar19} beta asymmetry ($A$-coefficient) experiments.
\par
Recoil-order and radiative corrections to equation \ref{E:JTWeqn} enter at about the 0.1\% level ({\em i.e.} an $\approx1$ \% relative contribution to the $a$-coefficient),  important at the precision level of recent and future experiments, especially for comparing the value of $\lambda$ obtained using different methods. The purpose of this paper is to describe and apply these corrections to previous results of the aCORN experiment \cite{Dar17, Has21} and make a comparison to other related experimental results.

\section{Recoil-order correction}
\label{S:Recoil}
The charged weak hadronic current between a neutron and proton contains six Lorentz-invariant terms (see for example ref.~\cite{Com83})
\begin{multline}
\label{E:weak}
\langle p(p_2) | V^{\mu} - A^{\mu} | n(p_1) \rangle = \overline{u}_p(p_2) [ f_1(q^2)\gamma^{\mu} + i\frac{f_2(q^2)}{M_n}\sigma^{\mu \nu}q_{\nu} + \frac{f_3(q^2)}{M_n}q^{\mu} \\
- g_1(q^2)\gamma^{\mu}\gamma_5 - i\frac{g_2(q^2)}{M_n}\sigma^{\mu \nu}\gamma_5 q_{\nu} - \frac{g_3(q^2)}{M_n}\gamma_5 q^{\mu} ] u_n(p_1),
\end{multline}
where $q = p_1 - p_2$ is the four-momentum transfer and $M_n$ is the neutron mass. The form factors $f_1$, $f_2$, $f_3$ are associated with the vector weak current $V^{\mu}$ with $f_1(q^2 \rightarrow 0) = G_V$. The recoil order $f_2$, $f_3$ terms are smaller by a factor of $q/M_n \approx 10^{-3}$. Similarly the form factors $g_1$, $g_2$, $g_3$ are associated with the axial vector weak current $A^{\mu}$ with $g_1(q^2 \rightarrow 0) = G_A$, and $g_2$, $g_3$ are recoil order. In the SM the ``second class'' currents $g_2$ and $f_3$ and the induced pseudoscalar form factor $g_3$ are known to be very small, so we will omit them in what follows. According to the Conserved Vector Current (CVC) hypothesis \cite{Fey58,Gel58} of the SM, the $f_2$ ``weak magnetism'' factor must equal the corresponding magnetic form factor in the electromagnetic current and so is given by the difference of the anomalous magnetic dipole moments of the proton and neutron. 
\par
The beta decay matrix element including the weak hadronic and leptonic currents is
\begin{equation}
{\mathcal M} = \frac{G_F}{\sqrt{2}}\langle p(p_2) | V^{\mu} - A^{\mu} | n(p_1) \rangle\,\big[\overline{u}_e(p_e)\gamma_{\mu}(1 + \gamma_5)u_{\nu}(p_{\nu}) \big].
\end{equation}
Harrington \cite{Har60} derived the beta decay differential decay rate
\begin{equation}
\label{E:Harrington}
d^3\Gamma = \frac{G_F^2}{2(2\pi)^5} \,   \frac{|{\bf p_e}||{\bf p_{\nu}}|}{M_n - E_e + |{\bf p_e}| \cos\theta_{e\nu}}  \Big[C_1 
+ {\bf P}\cdot \big( C_2 \mbox{$\bf{p_e}$} + C_3 \mbox{$\bf{p_{\nu}}$} + C_4 (\mbox{$\bf{p_e} \times \bf{p_{\nu}}$}) \big) \Big] dE_e d\Omega_e d\Omega_{\nu},
\end{equation}
where $G_F = 1.16638 \times 10^{-5}$ GeV$^{-2}$ is the Fermi weak coupling constant. The $C_i$ are complicated expressions, defined in Ref.~\cite{Har60}, that contain the form factors of equation \ref{E:weak} and  $\theta_{e\nu}$ is the angle between the electron and antineutrino momentum vectors in the decay frame. Equations \ref{E:JTWeqn} and \ref{E:Harrington} can be related by expanding equation \ref{E:Harrington} in $(q/M_n)$ and adding the first recoil order terms to the correlation coefficients in equation \ref{E:JTWeqn}, which then become functions of $E_e$. The result for the $a$-coefficient has been given by several authors \cite{Bil60, Nac68, Gar01}.

\begin{multline}
\label{E:aRecoil}
a(E_e,\theta_{e\nu},\lambda) = \frac{1 - \lambda^2}{1 + 3\lambda^2} + \frac{1}{(1 + 3\lambda^2)^2}\, \Big\{\frac{\epsilon}{Rx} (1-\lambda^2)(1 + 2| \lambda| + \lambda^2 + 4| \lambda |\tilde{f_2})
+ 4R (1+\lambda^2) (\lambda^2 + | \lambda | + 2| \lambda | \tilde{f_2})\\
 - Rx \big[ 3(1 + 3\lambda^2)^2 + 8| \lambda |(1+\lambda^2)(1 + 2\tilde{f_2}) + 3(1 - \lambda^2) (1 + 3\lambda^2) \beta \cos\theta_{e\nu} \big] \Big\}.
\end{multline}
Here $R$ = $E_e^{\rm max}$/$M_n$ = $1.37539\times 10^{-3}$, $\epsilon$ = $(m_e/M_n)^2$ = $2.95792\times 10^{-7}$, $x$ = $E_e$/$E_e^{\rm max}$, $\beta$ = $\sqrt{1 - (m_e/E_e)^2}$, 
$\tilde{f_2}$ = $f_2(0)$/$f_1(0)$ = 1.85295, and $m_e$ is the electron mass. The approximation $q^2 \approx 0$ is appropriate here because the $q^2$ dependence of the form factors enters at next-to-leading recoil order. 

\section{Radiative correction}
\label{S:Rad}
We will focus on the ``outer" radiative correction that affects the beta spectrum and decay correlations such as the $a$-coefficient. This correction accounts for virtual photons exchanged between final state particles (proton, electron) as well as real bremsstrahlung photons that are emitted. A recent paper \cite{Glu23} emphasized that the outer radiative correction for an $a$-coefficient measurement depends on the experiment. In particular a different result is obtained if the photon degrees of freedom are integrated while fixing the antineutrino momentum (referred to as ``neutrino type" in ref.~\cite{Glu23}) compared to the calculation where the proton momentum is fixed (``recoil type"). Both types can be found in the literature on beta decay radiative corrections, but in any practical neutron decay $a$-coefficient experiment the proton is detected, not the antineutrino, so the recoil type correction is needed.
\par
The presence of a real bremsstrahlung photon changes the final state kinematics from 3-body to 4-body. This is significant for experiments such as aCORN that determine the $a$-coefficient from a momentum correlation between the electron and recoil proton. Because the cold neutron effectively decays at rest and the antineutrino is not observed, we have $ {\bf p_e}\cdot{\bf p_{\nu}} = - {\bf p_e}\cdot({\bf p_e} + {\bf p_{\rm proton}})$. If a bremsstrahlung photon with momentum ${\bf k}$ is added to the final state this becomes $ {\bf p_e}\cdot{\bf p_{\nu}} = - {\bf p_e}\cdot({\bf p_e} + {\bf p_{\rm proton}} + {\bf k})$ and averaging over ${\bf k}$ makes the observed electron-antineutrino correlation differ from that implied by Eq. \ref{E:JTWeqn}. The result is very sensitive to details of the experiment so the radiative correction must be included in the experimental analysis; it cannot be simply applied after the fact.

\section{Corrections to the \lowercase{a}CORN result}
\label{S:aCORN}
\begin{figure}
\centering
\includegraphics[width = 4.0in]{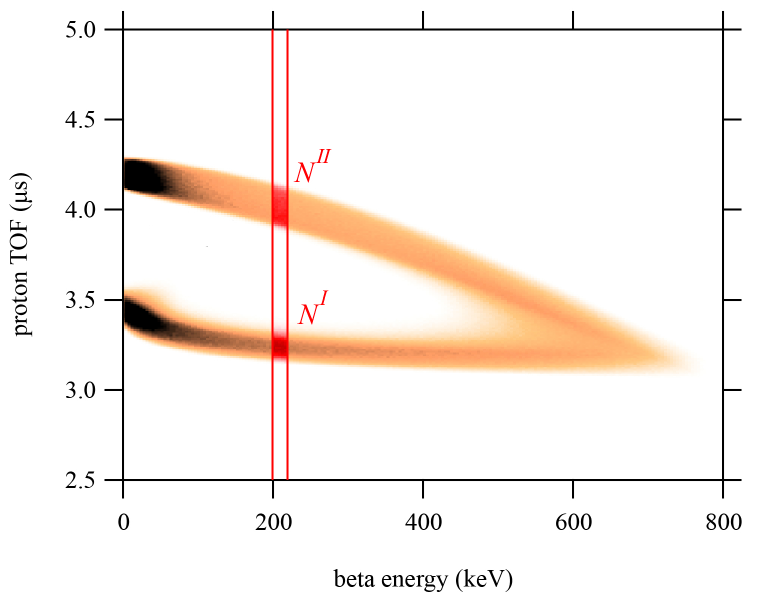}
\caption{\label{F:wishbone} A plot of proton TOF {\em vs.} electron energy for Monte Carlo aCORN data. The asymmetry in count rate of the upper and lower branches of the wishbone is used to determine the $a$-coefficient.}
\end{figure}

The aCORN neutron decay experiment ran on the NG-6 and NG-C cold neutron beam lines at the NIST Center for Neutron Research from 2013--2016. The collaboration published a combined result for the $a$-coefficient with uncertainty 1.7 \%  in 2021 \cite{Has21}. Details of the method, experimental apparatus, data analysis, and systematic effects can be found in previous publications \cite{Dar17, Has21, Col17, Has17, Sch21}. aCORN detects the beta electron and recoil proton from neutron decay in delayed coincidence. For each coincidence event the beta energy and proton time-of-flight (TOF) relative to electron detection are recorded and plotted as shown in figure \ref{F:wishbone}. The momentum acceptances created by the aCORN magnetic field and collimators cause these events to form a wishbone-shaped distribution with the lower (upper) branch containing decays where the electron and antineutrino were emitted into the same (opposite) hemisphere. For a vertical slice at each electron energy, the ``wishbone asymmetry" $X(E)$ of the lower ($N^I$) and upper ($N^{II}$) branch event rates is approximately proportional to the $a$-coefficient times a geometric function $f_a(E)$, a momentum space acceptance function that depends on the magnetic field shape and collimator geometry
\begin{equation}
\label{E:XE}
X(E) = \frac{N^I - N^{II}}{N^I + N^{II}} = a f_a(E) \left[ 1 + \delta_1(E) \right] + \delta_2(E)
\end{equation}
with
\begin{equation}
\label{E:faE}
f_a(E) = \frac{1}{2} v \left( \phi^I(E) - \phi^{II}(E) \right)
\end{equation}
and
\begin{equation}
\phi^I(E) = \frac{ \int d\Omega_e \int_I d\Omega_{\nu} \cos\theta_{e\nu} }{\Omega_e \Omega_{\nu}^I} \nonumber
\end{equation}
\begin{equation}
\label{E:phis}
\phi^{II}(E) = \frac{ \int d\Omega_e \int_{II} d\Omega_{\nu} \cos\theta_{e\nu} }{\Omega_e \Omega_{\nu}^{II}}
\end{equation}
and $v$ is the electron velocity in units of $c$.
Equations \ref{E:phis} can be understood as the average value of $\cos\theta_{e\nu}$ over the momentum acceptances for lower branch (I) and upper branch (II) events. They are computed as a function of electron energy by Monte Carlo integration. 
The small energy-dependent correction $\delta_1(E)$ is given by
\begin{equation}
\delta_1(E) = -\frac{1}{2} a v \left( \phi^I(E) + \phi^{II}(E) \right)
\end{equation}
and has a numerical value $<5\times 10^{-3}$. The other small energy-dependent correction $\delta_2(E)$ is the piece of the wishbone asymmetry $X(E)$ that is independent of the $a$-coefficient. It arises when the  proton kinetic energy, about 0.1 \% of the beta decay energy, is included. It also obtains a contribution from the momentum of inner bremsstrahlung photons in the radiative correction.
\par
Due to the electron and proton transverse momentum constraints in aCORN, analytical or semianalytical
calculation of the bremsstrahlung integral part of the radiative correction is very difficult, so instead we used a Monte Carlo method.
Many (order  $10^{13}$)  neutron decay events without and with bremsstrahlung photons were
generated and selected using the momentum constraints in a model of the aCORN experiment.
We used the recently written C++ Monte Carlo code called GENDER
(GEneration of polarized Neutron (and nuclear beta) Decay Events with
Radiative and recoil corrections) \cite{GluckGENDER}.
The Monte Carlo generation method is similar to the older FORTRAN code
that is described in ref. \cite{Glu97}:
\begin{enumerate}
\item Electron and neutrino direction vectors are generated for zeroth-order and virtual soft bremsstrahlung events with 3-body decay kinematics. Here the bremsstrahlung photon energy is very small.
\item Electron, neutrino and photon direction vectors are generated for hard bremsstrahlung (BR) events with 4-body decay kinematics.
\item Importance sampling is used in the case of hard BR generation.
\end{enumerate}
Nevertheless GENDER contains some improvements:  relativistic kinematics are exact, both in the soft and hard bremsstrahlung cases; and recoil-order corrections
are included in the zeroth-order calculations. In GENDER there are two possibilities (C++ class member functions) for the neutron decay event
generation: unweighted and weighted event generators. The unweighted event generator is easier to use but leads to larger statistical errors for the radiative corrections. On the other hand, the weighted event generator is somewhat more demanding to use but the radiative corrections have much smaller statistical errors. GENDER was thoroughly tested using various test computations such as comparisons with
analytical radiative correction results in the literature (see ref. \cite{Glu93} and references therein), and comparisons with various numerical radiative correction results \cite{Glu93}. The GENDER code results are also in good agreement with the analytical radiative correction results of reference \cite{Sen23}.
\begin{figure}
\centering
\includegraphics[width = 4.0in]{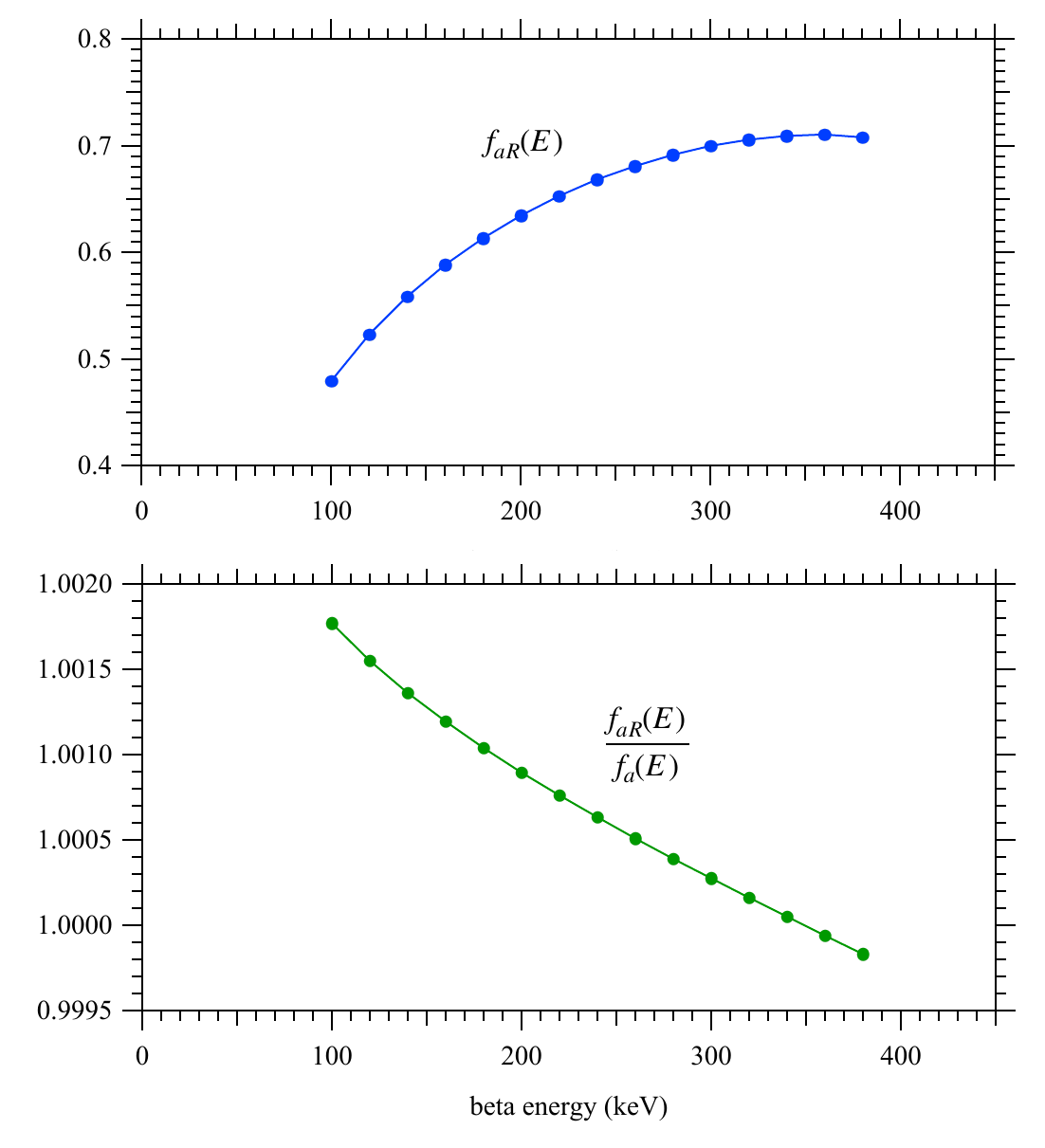}
\caption{\label{F:faRE} Top: The aCORN geometric function $f_{aR}(E)$, that includes the outer radiative correction. Bottom: The ratio of $f_{aR}(E)$ and the uncorrected $f_a(E)$.}
\end{figure}
\begin{figure}[h!]
\centering
\includegraphics[width = 4.0in]{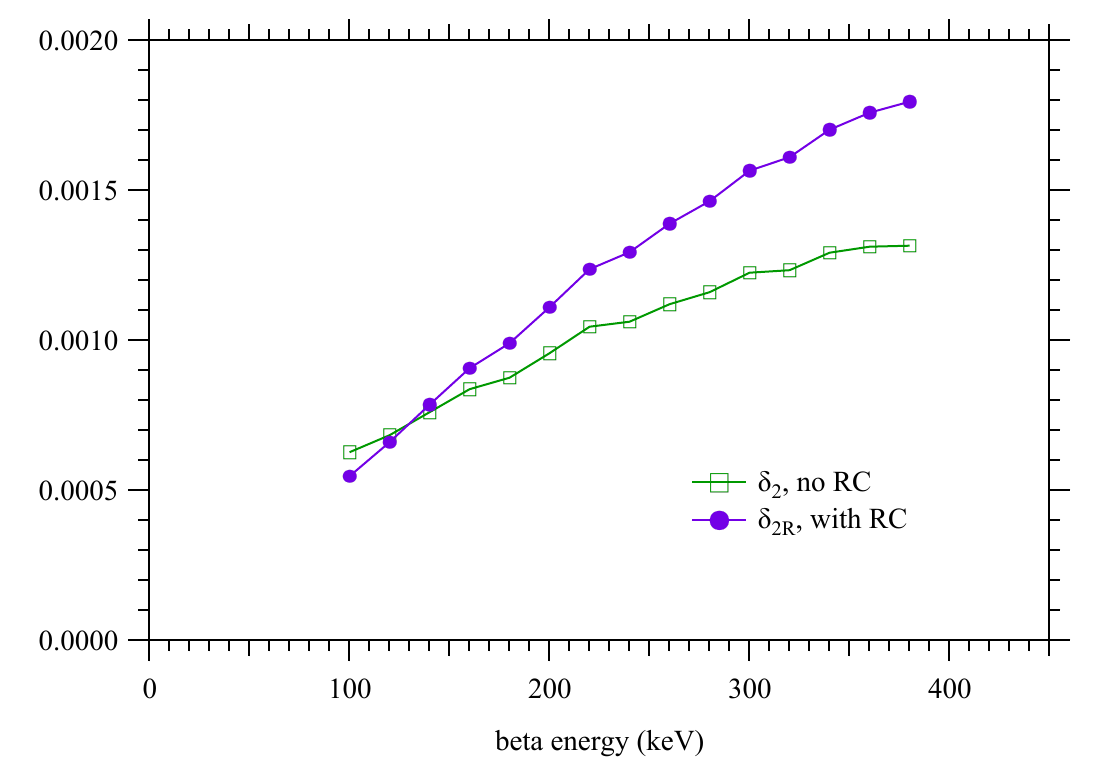}
\caption{\label{F:delta2} The aCORN correction function $\delta_2$ with and without the radiative correction.}
\end{figure}
\FloatBarrier
\par
Figures \ref{F:faRE} and \ref{F:delta2} show the results of our calculations for the order-$\alpha$ outer (model-independent) radiative-corrected aCORN functions $f_{aR}(E)$ and $\delta_{2R}(E)$ and comparisons to the uncorrected versions $f_{a}(E)$ and $\delta_{2}(E)$  for 15 electron energy bins (each bin has 20 keV width). For these computations we used the GENDER weighted event generator. Due to the presence of the bremsstrahlung photons and their influence on the decay kinematics, the
radiative correction results are generally sensitive to experimental details; this statement is especially
pertinent for the aCORN experiment.  
\begin{figure}
\centering
\includegraphics[width = 4.4in]{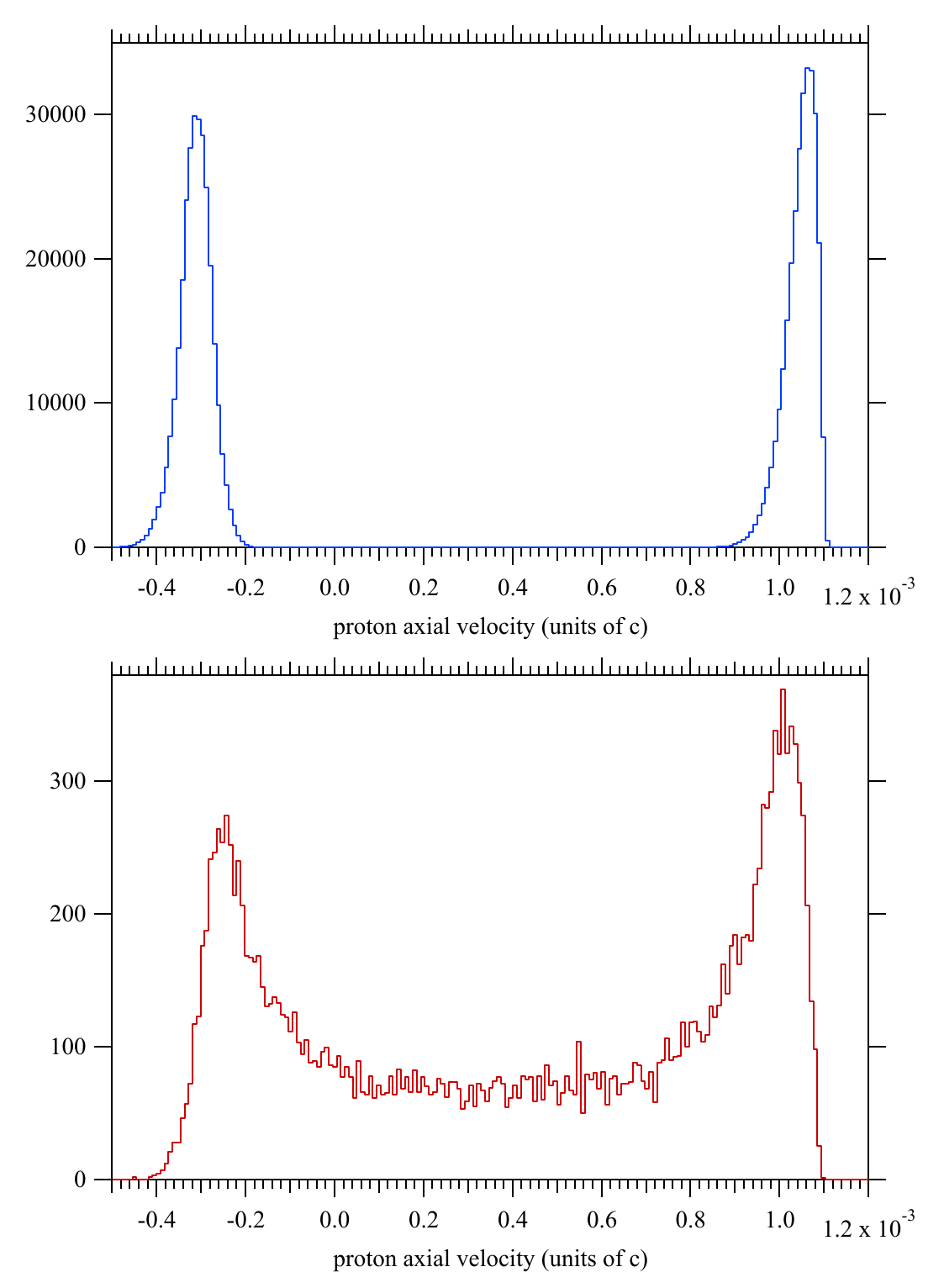}
\caption{\label{F:FG} Top: The Monte Carlo aCORN proton axial velocity distribution, without the radiative correction, for wishbone events with beta energy 110-130 keV. Bottom: 
The same distribution with the radiative correction, where a hard bremsstrahlung photon of energy $>$ 20 keV was emitted in the decay (note change of vertical scale).}
\end{figure}
\par
\begin{figure}
\centering
\includegraphics[width = 4.5in]{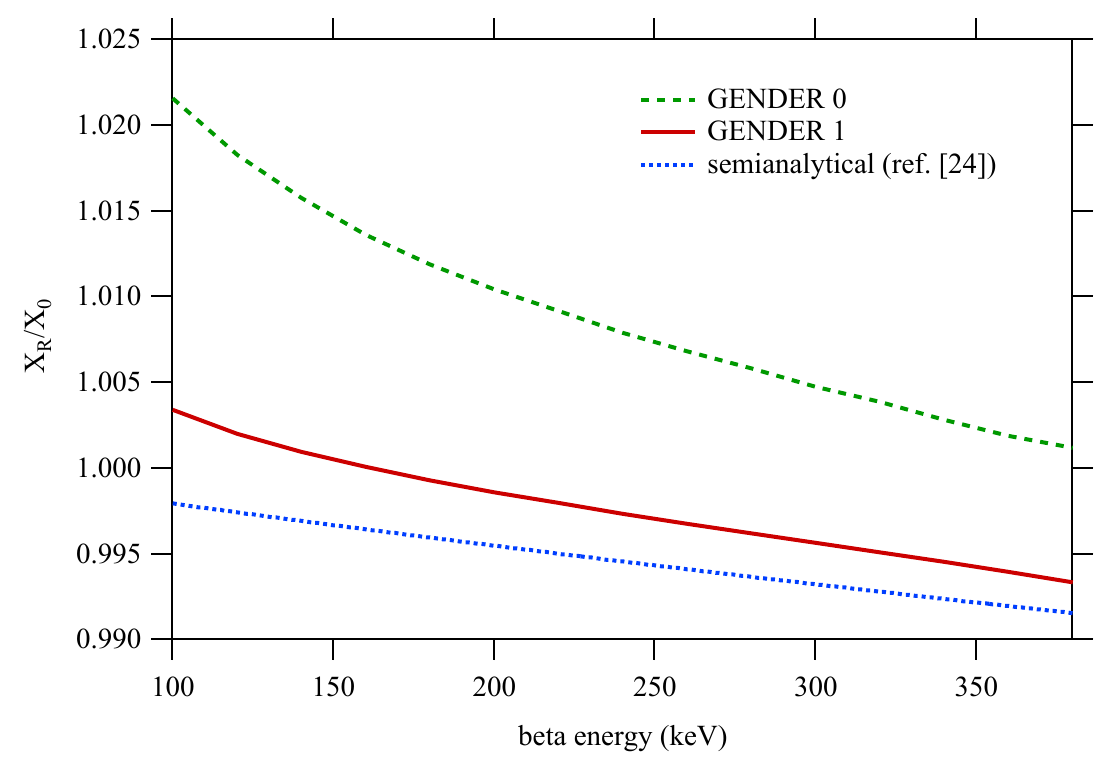}
\caption{\label{F:Xratio} The ratio of Monte Carlo wishbone asymmetries with the radiative correction $X_R$ and without $X_0$ calculated three different ways: using the GENDER code and separating the wishbone branches simply by the average proton energy (GENDER 0); using the GENDER code and separating the wishbone branches at the same points used in the aCORN analysis (GENDER 1, the correct method); and using the semianalytical formulation in reference \cite{Glu93} which does not consider the proton energy acceptance of the experiment.}
\end{figure}

Figure \ref{F:FG} shows the axial velocity distribution of 
neutron decay protons for the 110--130 keV electron energy bin for two cases: the zeroth-order
calculation without bremsstrahlung photons; and  when a bremsstrahlung photon \mbox{$>$20} keV was emitted.
One can see that without photons the low and high velocity peaks are clearly separated, but with higher energy 
photons there are events also between the peaks. Therefore both the sign and magnitude of the radiative correction to aCORN depends sensitively on how the wishbone branches are separated in the analysis and care must be taken to do this correctly. To illustrate this point, figure \ref{F:Xratio} shows three different results for the ratio of Monte Carlo wishbone asymmetries $X_R$ (with radiative correction) and $X_0$ (without). For the green dashed line labeled ``GENDER 0'' the wishbone branches were separated simply by the average proton energy. For the red solid line labeled ``GENDER 1'' the wishbone branches were separated at the same points used in the aCORN analysis, which were optimized to minimize the systematic error. This is the result appropriate for the aCORN correction. As a comparison, the blue dotted line shows the result using the semianalytical formulation in reference \cite{Glu93}, Section 4 (table 5). In that calculation, the electron energy and the experimentally defined electron-neutrino
angle (eq. 4.1 in reference \cite{Glu93}) are fixed and the proton energy is integrated out during the bremsstrahlung photon phase space integration part of the radiative correction calculation. Therefore, although having some kind of similarity with the aCORN experimental details, the semianalytical calculation cannot be applied for the analysis of the aCORN experiment where the proton's transverse energy is constrained.

\par
\begin{figure}
\centering
\includegraphics[width = 4.5in]{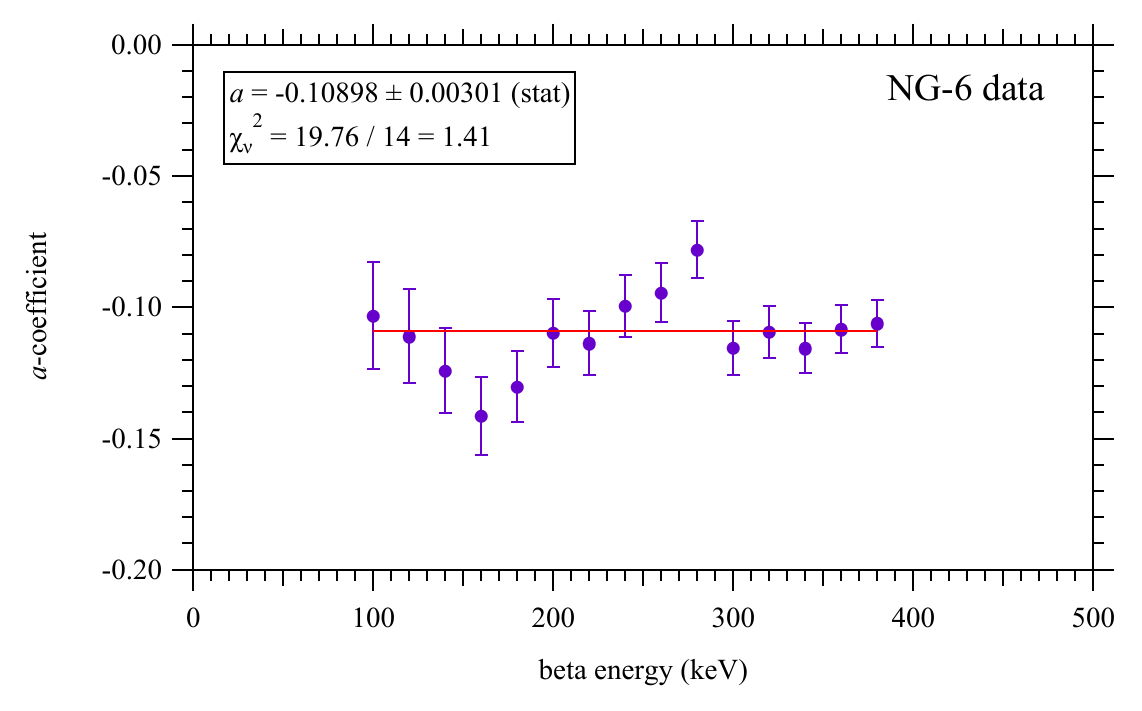}
\caption{\label{F:NG6aRad} A weighted average of the experimental $a$-coefficient data, including the outer radiative correction, from the NG-6 run.}
\end{figure}
\begin{figure}
\centering
\includegraphics[width = 4.5in]{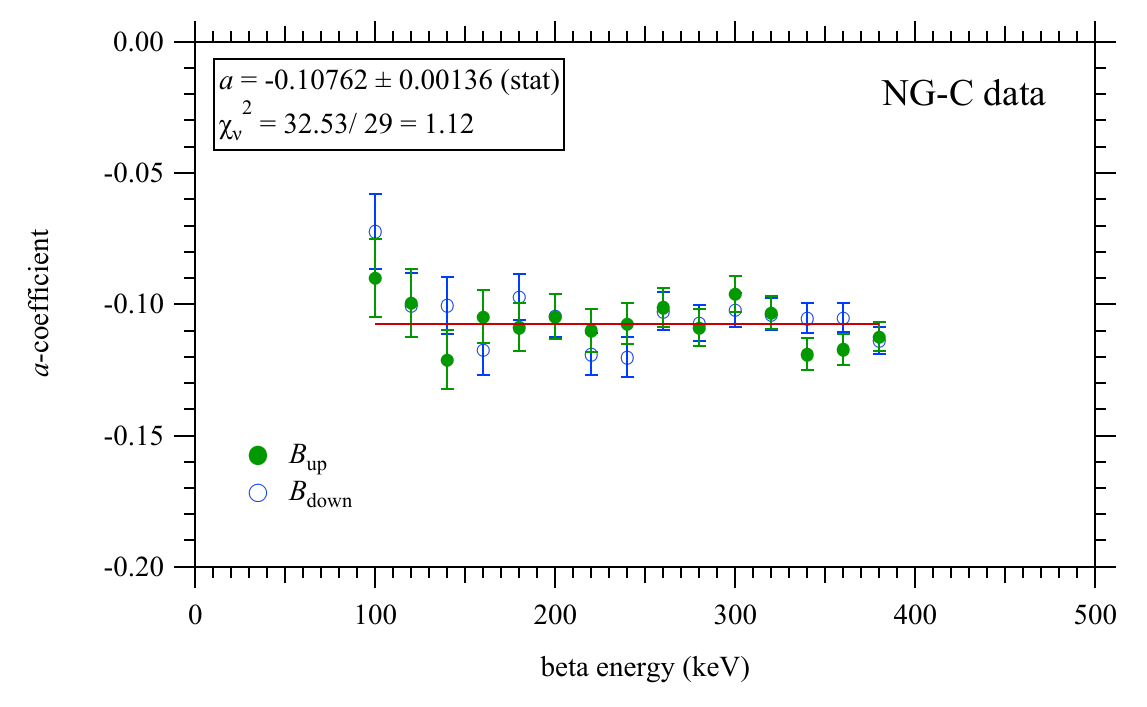}
\caption{\label{F:NGCaRad} A weighted average of the experimental $a$-coefficient data, including the outer radiative correction, from the NG-C run.}
\end{figure}

With the radiative-corrected functions $f_{aR}(E)$ and $\delta_{2R}(E)$ in hand, the data are analyzed in the same manner as described in reference \cite{Has21}. The experimental result for the 
$a$-coefficient is found from the wishbone asymmetry in each energy bin with
\begin{equation}
a(E) = \frac{X(E) - \delta_{2R}(E)}{  f_{aR}(E) \left[ 1 + \delta_1(E) \right] }
\end{equation}
The result $a(E)$ for the NG-6 data is shown in figure \ref{F:NG6aRad}. Here the simple average of $X(E)$ for the magnetic field up ($B_{\rm up}$) and down ($B_{\rm down}$) data was taken to correct for the observed difference we attributed to residual neutron polarization (see the discussion in ref.~\cite{Dar17}). These were fit to a constant to obtain the weighted average of the $a$-coefficient. Figure \ref{F:NGCaRad} presents a similar analysis of the NG-C data, except the $B_{\rm up}$ and $B_{\rm down}$ data are shown separately and fit together, as the residual polarization on NG-C was found to be negligible \cite{Sch21}. Combining the NG-6 and NG-C results, including the systematic uncertainties discussed in Ref.~\cite{Dar17, Has21}, we obtain $a = -0.10785 \pm .00125{\rm (stat)} \pm .00133{\rm (sys)}$ which is essentially unchanged from the combined aCORN result reported in ref.~\cite{Has21}. While one may expect {\em a priori} to see a shift in the $a$-coefficient at the level of $\approx$1 \% due to the radiative correction, both the direction and magnitude of the shift depend sensitively on details of the experiment. In the case of aCORN the corrections to $f_{a}(E)$ and $\delta_{2}(E)$ are small and work in opposite directions. We also discovered and corrected a small inaccuracy in the method of calculating $\delta_{2}(E)$ in reference \cite{Has21} that causes a 0.2 \% change in the result. The combination of all these happen to effectively cancel.

\par
\begin{figure}
\centering
\includegraphics[width = 4.5in]{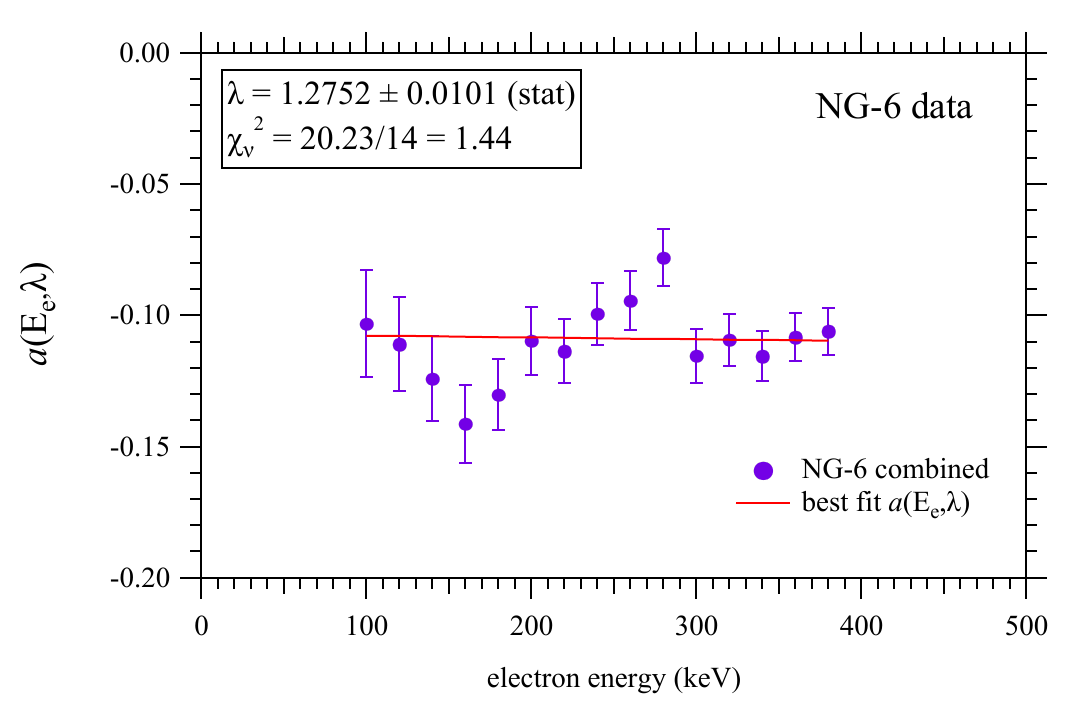}
\caption{\label{F:NG6RadRec} A fit of the experimental $a$-coefficient data from the NG-6 run, including the outer radiative correction, to the recoil corrected function $a(E_e, \lambda)$ of Eq. \ref{E:aRecoil}, varying $\lambda$ as a free parameter.}
\end{figure}
\begin{figure}
\centering
\includegraphics[width = 4.5in]{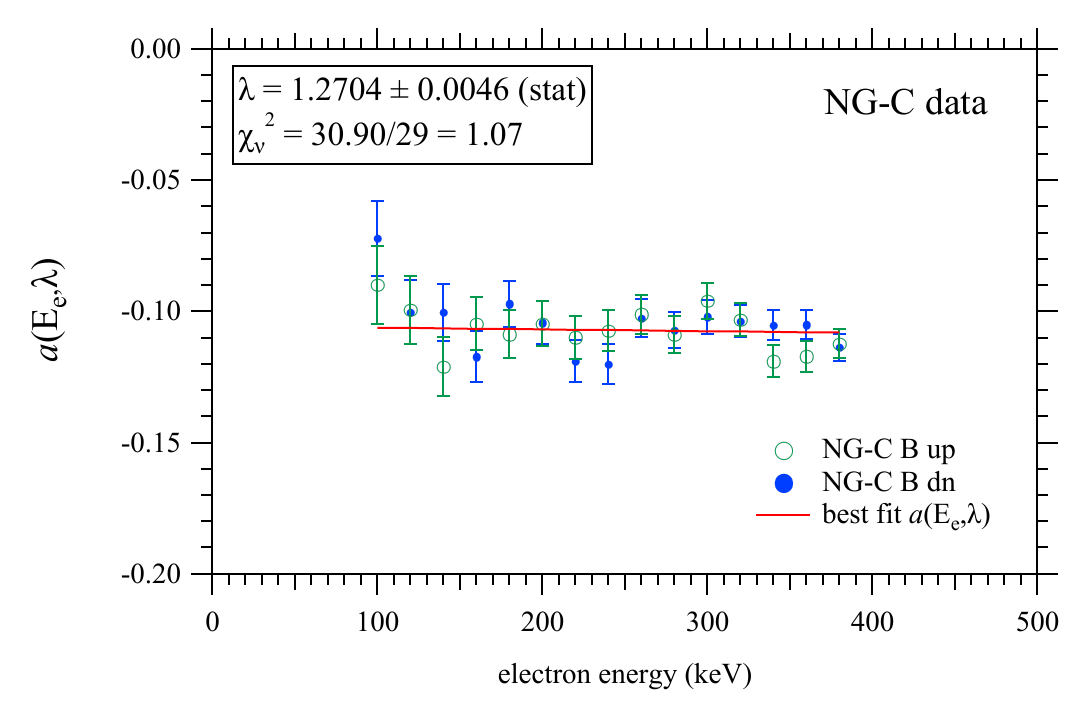}
\caption{\label{F:NGCRadRec} A fit of the experimental $a$-coefficient data from the NG-C run, including the outer radiative correction, to the recoil corrected function $a(E_e, \lambda)$ of Eq. \ref{E:aRecoil}, varying $\lambda$ as a free parameter.}
\end{figure}

To include the first recoil order correction, we fit the experimental $a$-coefficient data in figures \ref{F:NG6aRad}, \ref{F:NGCaRad} to the energy-dependent function $a(E_e,\theta_{e\nu},\lambda)$ given by Eq. \ref{E:aRecoil}, varying $\lambda$ as a free parameter to minimize chi-squared. In equation \ref{E:JTWeqn} we are concerned with the decay correlation that is linear in $\cos\theta_{e\nu}$, so it is appropriate to average over the domain of $\cos\theta_{e\nu}$ in equation \ref{E:aRecoil} to remove the quadratic dependence in the last term. For all neutron decays, and in most previous experiments, this averages to zero. For aCORN, due to the limited momentum acceptances for detecting protons and electrons, the average is nonzero and energy dependent. This average, calculated by Monte Carlo, has been included but it is negligibly small: the last term in equation \ref{E:aRecoil} makes a relative contribution of 0.3\% to the full recoil correction.
The results of these fits are shown in figures \ref{F:NG6RadRec} and \ref{F:NGCRadRec}. They are in good agreement so we combine them to give an overall result for $\lambda$, including the systematic uncertainties described in ref.~\cite{Dar17,Has21}: 
\begin{equation}
\lambda = -1.2712 \pm 0.0061.
\end{equation}

With this result for $\lambda$ we can update our combined result for the $a$-coefficient using Eq. \ref{E:aRecoil} with a weighted average over the aCORN energy range 100 keV -- 380 keV
\begin{equation}
\label{E:aCORNres}
\langle a \rangle = -0.10785 \pm 0.00125\, ({\rm stat}) \pm 0.00133\, ({\rm sys}). 
\end{equation}
An alternative way to present this result, that may be generally more useful given the energy dependence at recoil order, is the extrapolated average value of the $a$-coefficient weighted by the full theoretical Fermi beta spectrum for neutron decay, for which we introduce the notation $\overline{a}$
\begin{equation}
\label{E:aBar}
\overline{a} = -0.10779 \pm 0.00125\, ({\rm stat}) \pm 0.00133\, ({\rm sys}). 
\end{equation}
We note that adding the recoil effect had a very small effect on the value of the $a$-coefficient because the energy dependence is relatively weak, but the resulting value of $\lambda$ due to the combined radiative and recoil corrections shifts by \mbox{0.66 \%} relative to that reported in Ref.~\cite{Has21}.\\
\FloatBarrier
\section{Comparison to previous experiments}
Previous experiments since the 1970s obtained the $a$-coefficient from the shape of the neutron decay recoil proton energy spectrum. Before making a comparison we will briefly review what those experiments measured and how they differ from aCORN. The theoretical proton recoil spectrum, correct to first recoil order, was derived by Pietschmann \cite{Pie68} and Nachtmann \cite{Nac68}
\begin{equation}
\label{E:Piet}
N(T) = \frac{G^2_V\,M_n}{4 \pi^3} \int\limits_{E_e^{(1)}}^{E_e^{(2)}} F(E_e - m_e) \Big\{ 2(\Delta - E_e) E_e (1 + \lambda^2)
 - \frac{1}{2} (\Delta^2 - m^2_e) (1 - \lambda^2) + (1 - \lambda^2) M_n T \Big\} dE_e
\end{equation}
where
\begin{equation}
E_e^{(1,2)} = \frac{ (\Delta \mp \sqrt{2 M_p T})^2 + m^2_e}{2(\Delta \mp \sqrt{2 M_p T})}
\end{equation}
are the limits of the kinematically allowed range of $E_e$ for a given value of the proton kinetic energy $T$.
Here $\Delta = M_n - M_p$ is the neutron-proton mass difference. The Fermi function $F(E_e - m_e)$ includes Coulomb corrections. Note that equation \ref{E:Piet} depends on $\lambda^2$ but not directly on the neutron $a$-coefficient of equation \ref{E:JTWeqn}. Now defining \cite{Byr02}
\begin{equation}
\label{E:a0}
a_0 \equiv \frac{1 - \lambda^2}{1 + 3\lambda^2},
\end{equation}
equation \ref{E:Piet}  can be rewritten as \cite{Str78}
\begin{multline}
\label{E:Strat}
N(T) = \frac{G^2_V\,M_n}{4 \pi^3} \frac{1}{1 + 3 a_0}  \Big\{ \int\limits_{E_e^{(1)}}^{E_e^{(2)}} F(E_e - m_e) 4(\Delta - E_e) E_e\,dE_e\\
+ a_0 \int\limits_{E_e^{(1)}}^{E_e^{(2)}} F(E_e - m_e) [4(\Delta - E_e) E_e - 2(\Delta^2 - m^2_e) + 4 M_n T] \,dE_e \Big\}.
\end{multline}
Previous experiments that obtained the neutron $a$-coefficient from the recoil proton spectrum \cite{Str78,Byr02,Bec20} measured $a_0$ as defined by equations \ref{E:a0}, \ref{E:Strat} (complete to first recoil order per equation \ref{E:Piet}). Experiments that obtain the $a$-coefficient from the angular correlation of beta electrons and recoil protons, such as aCORN and the currently-running Nab experiment \cite{Fry19} measure $a(E_e,\theta_{e\nu},\lambda)$ given by equation \ref{E:aRecoil} to first recoil order. These are not the same, but are equivalent functions of $\lambda^2$ when recoil order terms are omitted. In particular $a(E_e,\theta_{e\nu},\lambda)$ is a function of electron energy and depends in first recoil order on the weak magnetism form factor $f_2$, while $a_0$ does not. This distinction has often been disregarded in the previous literature where ``$a$'' was typically used to refer both. Here we will use $a_0$ to refer to the expression in equation \ref{E:a0}, and $\overline{a}$ to refer to the result measured by aCORN averaged over the Fermi beta spectrum and $\beta \cos\theta_{e\nu}$ as described in section \ref{S:aCORN}. We emphasize in this context that $a_0$ and $\overline{a}$, while different, are both complete to first recoil order.
\begin{table}[b]
\caption{\label{T:lambda} A summary of results for $a_0$ and $\lambda$ from neutron $a$-coefficient experiments. In the case of aCORN, $a_0$ is calculated from $\lambda$ 
using equation \ref{E:a0}. }
\centering
\begin{ruledtabular}
\begin{tabular}{lllllll}
Experiment & Year & Ref.  & method & $a_0$ & $\lambda$  \\ \hline
Stratowa, {\em et al.} & 1978 & \cite{Str78} & $p$ spectrum & -0.1017 $\pm$ 0.0051 & -1.259 $\pm$ 0.017\\
Byrne, {\em et al.} & 2002 & \cite{Byr02} & $p$ spectrum & -0.1054 $\pm$ 0.0055 & -1.271 $\pm$ 0.019\\
aSPECT & 2020 & \cite{Bec20} & $p$ spectrum & -0.10430 $\pm$ 0.00084 & -1.2678 $\pm$ 0.0028\\
aCORN & 2023 & this work & asymmetry & -0.1053 $\pm$ 0.0018 & -1.2712 $\pm$ 0.0061\\
\end{tabular}
\end{ruledtabular}
\end{table}
\begin{figure}
\centering
\includegraphics[width = 4.5in]{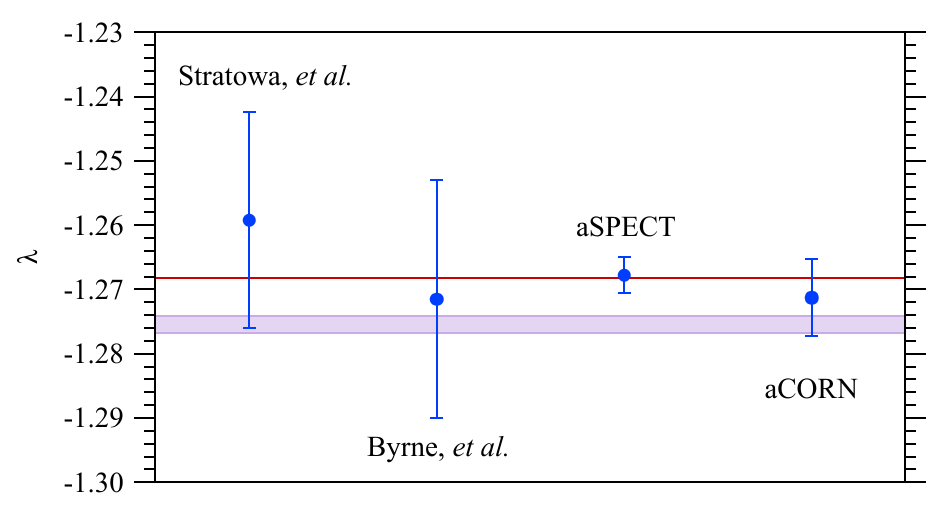}
\caption{\label{F:aSummary} A summary of the results for $\lambda = G_A / G_V$ from neutron $a$-coefficient experiments. The horizontal line is the weighted average:
$\lambda = -1.2686 \pm 0.0025$. The shaded region shows the 2022 PDG recommended value $\lambda = -1.2754 \pm 0.0013$.}
\end{figure}

\par
In light of this it is not meaningful to directly compare the aCORN result for $\overline{a}$ to $a_0$ measured by the proton recoil spectrum experiments.  It is better to compare the results for $\lambda$, or equivalently $a_0$, determined by all the experiments. This comparison is shown in table \ref{T:lambda}, and for $\lambda$ in figure \ref{F:aSummary}. The aCORN result for $a_0$, calculated from $\lambda$ 
using equation \ref{E:a0}, differs from the $\overline{a}$ result (equation \ref{E:aBar}) by 2.4 \%. Results from these four experiments are in good agreement and yield the weighted average $\lambda = -1.2682 \pm 0.0025$. There is some tension between this and the Particle Data Group (PDG) 2022 \cite{PDG22} recommended value $\lambda = -1.2754 \pm 0.0013$ (expanded uncertainty).  The $a$-coefficient average disagrees with the PERKEO III beta asymmetry result \cite{Mar19}: $\lambda = -1.27641 \pm 0.00056$ by more than 3$\sigma$, a discrepancy driven mainly by the aSPECT result \cite{Bec20}.

\FloatBarrier
\section{Acknowledgements}
We thank Stefan Bae{\ss}ler and Susan Gardner for helpful discussions. This work was supported by the National Institute of Standards and Technology (NIST), U.S. Department of Commerce; National Science Foundation grants PHY-2012395 and PHY-1714461; and U.S. Department of Energy, Office of Nuclear Physics Interagency Agreement 89243019SSC000025. We acknowledge support from the NIST Center for Neutron Research, US Department of Commerce, in providing the neutron facilities used in this work.


\begin{thebibliography}{99}
\bibitem{JTW} J.~D.~ Jackson, S.~B.~Treiman, and H.~W.~Wyld, Nuclear Physics {\bf 4}, 206 (1957).
\bibitem{Dub91} D.~Dubbers, Nucl. Phys. {\bf A527}, 239 (1991). 
\bibitem{Bar05} J.~Barranco, G.~Miranda, and T.~I.~Rashba, JHEP {\bf 12}, 021 (2005).
\bibitem{Cyb16} R.~H.~Cyburt, B.~D.~Fields, K.~A.~Olive, and T-H~ Yeh, Rev.~Mod.~Phys. {\bf 88}, 15004 (2016).
\bibitem{Dub21} D.~Dubbers and B.~M\"{a}rkisch, Annu. Rev. Nucl. Part. Sci. 2021.71: 139--163 (2021).
\bibitem{Har15} J.~C.~Hardy and I.~S.~Towner, Phys. Rev. C {\bf 102}, 045501 (2020).
\bibitem{Bro18} M.~Brown, {\em et al.}, Phys.~Rev.~C {\bf 97}, 035505 (2018).
\bibitem{Mar19} B.~M\"{a}rkisch, {\em et al.}, Phys.~Rev.~Lett. {\bf 122}, 242501 (2019).
\bibitem{Bec20} M.~Beck, {\em et al.}, Phys.~Rev.~C {\bf 101}, 055506 (2020).
\bibitem{Dar17} G.~Darius, {\em et al.}, Phys. Rev. Lett. {\bf 119}, 042502 (2017).
\bibitem{Has21} M.~T.~Hassan, {\em et al.}, Phys. Rev. C {\bf 103}, 045502 (2021).
\bibitem{Com83} E.~D.~Commins and P.~H.~Bucksbaum, {\em Weak interactions of leptons and quarks}, Chapter 4, Cambridge University Press 1983, ISBN 0-521-27370-6.
\bibitem{Fey58} R.~P.~Feynman and M.~Gell Mann, Phys. Rev. {\bf 109}, 193 (1958).
\bibitem{Gel58} M.~Gell Mann, Phys. Rev. {\bf 111}, 362 (1958).
\bibitem{Har60} D.~R.~Harrington, {\em et al.}, Phys. Rev. {\bf 120}, 1482 (1960).
\bibitem{Bil60} S.~M.~Bilen'kii, {\em et al.}, JETP {\bf 37}, 1241 (1960).
\bibitem{Nac68} O.~Nachtmann, Zeit. f\"{u}r Phys. {\bf 215}, 505 (1968). In this paper a separate coefficient $b$ (not to be confused with the Fierz interference $b$) is used for the $\cos^2\theta$ term.
\bibitem{Gar01} S.~Gardner and C.~Zhang, Phys. Rev. Lett. {\bf 86}, 5666 (2001). Note: the difference in the last term between our expression and this reference is due to how the recoil-order $\cos^2\theta$ term is treated. The result in this reference cannot be directly applied to the aCORN analysis, although the difference is negligibly small.
\bibitem{Glu23} F.~Gl\"{u}ck, J. High Energy Phys. {\bf 9}, 188 (2023).
\bibitem{Col17} B.~Collett, {\em et al.}, Rev.~Sci.~Instr. {\bf 88}, 083503 (2017).
\bibitem{Has17} T.~Hassan, {\em et al.}, Nucl.~Instr.~Meth. A {\bf 867}, 51 (2017).
\bibitem{Sch21} B.~C.~Schafer, {\em et al.}, Nucl.~Instr.~Meth. A {\bf 988}, 1648662 (2021).
\bibitem{GluckGENDER} F. Gl\"uck, GENDER: GEneration of polarized Neutron (and nuclear beta) Decay Events with
Radiative and recoil corrections (to be published).
\bibitem{Glu97} F.~Gl\"{u}ck, Comp. Phys. Comm. {\bf 101}, 223 (1997).
\bibitem{Glu93} F.~Gl\"{u}ck, Phys. Rev. D {\bf 47}, 2840 (1993).
\bibitem{Sen23} C.-Y.~Seng, arXiv:2312.08630v1 (2023).
\bibitem{Pie68} H.~Pietschmann. Acta Phys. Austriaca, Suppl. (1968).
\bibitem{Byr02} J.~Byrne, {\em et al.}, J. Phys. G: Nucl. Part. Phys. {\bf 28}, 1325 (2002).
\bibitem{Str78} C.~Stratowa, R.~Dobrozemsky, and P.~Weinzierl, Phys. Rev. D {\bf 18}, 3970 (1978).
\bibitem{Fry19} J.~Fry, {\em et al.}, EPJ Web Conf. {\bf 219}, 04002 (2019).
\bibitem{PDG22} R.~L.~Workman, {\em et al.} (Particle Data Group), Prog. Theor. Exp. Phys. 2022, 083C01 (2022).

\end{thebibliography}
\end{document}